\newtheorem{theor}{Theorem}
\newtheorem{lemma}{Lemma}
\newcommand{\beq}{\begin{equation}}
\newcommand{\enq}{\end{equation}}
\newcommand{\bee}{\begin{eqnarray}}
\newcommand{\ene}{\end{eqnarray}}
\newcommand{\bem}{\begin{mathletters}}
\newcommand{\enm}{\end{mathletters}}

\newcommand{\crr}{{I\kern-.3468em R}}
\newcommand{\crz}{{Z\!\!\!\!I}}
\newcommand{\cx}{\sqcap\kern-.7000em \sqcup}
\newcommand{\fud}{{\frac{1}{2}}}
\newcommand{\fdt}{{\frac{d}{dt}}}
\newcommand{\cah}{{\cal H}}
\newcommand{\sig}{\sigma}

\newcommand{\be}{\beta}
\newcommand{\al}{\alpha}
\newcommand{\ga}{\gamma}
\newcommand{\de}{\delta}
\newcommand{\De}{\Delta}

\newcommand{\p}{\partial}

\newcommand{\du}{{\dot u}}
\newcommand{\ddal}{{\ddot\alpha}}
\newcommand{\dal}{{\dot\alpha}}
\newcommand{\ddbe}{{\ddot\beta}}
\newcommand{\dbe}{{\dot\beta}}

\newcommand{\intto}{\int^{t}_{0}}

\newcommand{\wta}{\tilde{a}}
\newcommand{\wtb}{\tilde{b}}
\newcommand{\wtu}{\tilde{u}}
\newcommand{\wtal}{\widetilde{\alpha}}

\newcommand{\wtpm}{\widetilde{P}}

\newcommand{\lV}{\left\Vert}
\newcommand{\rV}{\right\Vert}

\documentclass[11pt]{article}
\begin{document}

\begin{titlepage}
\thispagestyle{empty}

\title{ON GLOBAL EXISTENCE OF LOCALIZED SOLUTIONS
OF SOME NONLINEAR LATTICES}

\author{G.~Perla Menzala$^1$\footnote{E-mail: perla@lncc.br.}
and V. V. Konotop$^2$\footnote{E-mail: konotop@alf1.cii.fc.ul.pt}
\\ \\
$^1$National Laboratory of Scientific Computation, \\
Av. Getulio Vargas 333, 25651-070 Petropolis,  C.P.95113,  RJ Brazil
\\ and \\
Institute of Mathematics, UFRJ, RJ, Brazil
 \\
 \\
$^2$Centro de F\'{\i}sisca da Mat\'eria Condensada, \\
Complexo Interdisciplinar da Universidade de Lisboa, \\ Av.
Prof. Gama Pinto, 2, Lisboa, P-1649-003   Portugal \\ and \\
Departamento de F\'{\i}sica, Faculdade de Ci\^encias, Universidade de
Lisboa,\\ Campo Grande, 1749-016, Lisboa, Portugal}

\maketitle

\begin{center}
ABSTRACT
\end{center}

We prove global existence and uniqueness of solutions of some
important nonlinear lattices which include the Fermi-Pasta-Ulam (FPU)
lattice.  Our result shows (on a particular example) that the
FPU lattice with high nonlinearity and its continuum
limit display drastically different behaviour with respect to blow up
phenomenon.

\vspace{0.5 true cm}

\end{titlepage}

\section{Introduction}

In recent years nonlinear differential-difference equations, also called
lattices, attracted a great deal of attention. Various
aspects of spatially localized solutions of classical and quantum lattices
with high nonlinearity (i.e.~nonlinearity characterized by powers greater
than four) were discussed in a number of recent papers
(see e.g. \cite{BRC1,BRC2,Page} and references therein).
Related questions are of traditional interest in nonlinear physics
because on the one hand lattices serve as mathematical models for many
physical and biological systems like, for example, coupled oscillators,
macromolecules, spin chains, etc. On the other hand, the
discreteness introduces in evolution many new qualitative features compared
with the respective continuum limits.  Among them we can mention Bloch
oscillations of bright \cite{BLR,KCV} and dark \cite{K1}
solitons in the so-called Ablowitz-Ladik
model with a linear potential, dynamical localization of a soliton in a
periodic potential \cite{KCV} sensitivity of the lattice
dynamics to the type of the nonlinearity \cite{K2}.
One of the most important questions in the lattice dynamics is the
{\it existence} of various types of solutions.  An important step in
this direction has recently been made in [1, 15] where the proof of
existence and stability of breathers, i.e. solutions periodic in time
and localized in space, has been reported.  Those papers dealt with a
special limit, called anticontinuous, when the coupling between two
neighbour sites is small enough (i.e. when on-site potentials
dominate inter-site interactions).  Meanwhile the existence and stability
of localized excitations in more relevant physical situations when the
coupling in nonlinear lattices is of a general type remains an open
question (unlike the behaviour of various nonlinear continuum models,
see e. g. \cite{BBM,BPM,BS}). If the model considered appears with
some kind of dissipative mechanism once the global existence it is
settle then questions especially important for physical applications
are the stability and possible uniform rates of decay of the solutions
at time approaches infinity. Results in this direction will appear in
a forthcoming work \cite{PerKon}.

The purpose of the present paper is to prove global existence (and
uniqueness) of a solution of an important nonlinear lattice, the
so-called Fermi-Pasta-Ulam (FPU) lattice. In order to obtain our
results we built a convenient Banach space and get local solutions via
a contraction.  The calculations are lengthly because we work with
estimates of the explicit Green function associated to the problem.
We prefer it in this way because it becames clear in the proofs that
the techniques we use work as well for many other
important models with suitable modifications (see section 4).
The lattices are assumed to be described by Hamiltonians
of the following type
\beq
\label{e1}
\cah = \cah_\ell + \cah_{n\ell}
\enq
where
\beq
\label{e2}
\cah_\ell = \frac{M}{2} \sum^\infty_{n=-\infty} \du^2_n +
\frac{K_2}{2} \sum^\infty_{n=-\infty} \left ( u_{n+1} - u_n \right)^2,
\enq
with $u_n \equiv u_n (t)$ being a displacement of the $n$th particle,
$M>0$ being the mass of a particle, $K_2$ being a force constant
and $\dot{u}_n$ denotes the derivative with respect to time.
$\cah_\ell$ is the linear part describing nearest neighbour
interactions.  The nonlinear part $\cah_{n\ell}$ we will consider is
given by
\beq
\label{e3}
\cah_{n\ell} = \frac{K_{p+1}}{p+1} \sum^\infty_{n=-\infty}
\left ( u_{n+1} - u_n \right )^{p+1}.
\enq
where  $p$ is an integer, $p\ge 2$ and $K_{p+1}$ is a positive constant.

The paper is organized as follows: In Sections~2 and 3, we give
detailed proofs of local and global existence of spatially
localized solutions of the FPU lattice.  In Section~4 we present some
other important lattices for which the techniques apply as well with
suitable modifications, for example the "sine-Gordon" inter-site
nonlinearity where the nonlinear part is given by
\beq
\label{SG}
\cah_{n\ell} = \cah_{SG-IS} = K \sum^\infty_{n=-\infty} \left [
1 - \cos (u_{n+1} - u_n) \right ]
\enq
and the so-called on-site nonlinearity
\beq
\label{onsite}
\cah_{n\ell} = \cah_{OS} = \frac{K_{p+1}}{p+1} \sum^\infty_{n=-\infty}
u_n^{p+1}
\enq

\section{Local solutions of the FPU lattice}

The evolution equation for the FPU lattice reads
\bee
\label{2_1}
M\ddot{u}_n=K_2(u_{n+1}+u_{n-1}-2u_n)-K_{p+1}\left[\left(u_n-u_{n-1}\right)^p
+\left(u_n-u_{n+1}\right)^p\right]
\ene
All considerations below are restricted to the case when $p$ is odd
and $M$, $K_2$, and $K_{p+1}$ are positive constants.

After renormalization $\sqrt{K_2/M}t\; \mapsto t$ the equation of the
motion of the lattice (\ref{e1})-(\ref{e3}) is written in the form
\beq
\label{2_2}
\ddal_n (t) - \De\al_n (t) - \chi_p \De \al^p_n (t) = 0
\enq
where $n\in {\crz}$, $\al_n (t) = u_{n+1} - u_n$ is a relative displacement,
$\chi_p = K_{p+1}/K_2$, and $\De f_n \equiv f_{n+1} + f_{n-1} - 2f_n$.
We notice that introducing new variable $\al_n (t)$ makes not only
physical sense allowing to include kink-like solutions in the consideration.
The energy in the linearized problem for $\al_n (t)$ can be used as a
norm (see below) while the energy in the linearized problem for
$u_n (t)$ can be treated only as a seminorm.

We consider a Cauchy problem for equation (\ref{2_2}) subject to the
initial conditions
\beq
\label{2_3}
\al_n (t=0) = \wta_n = a_{n+1} - a_n;\qquad
\dal_n (t=0) = \wtb_n = b_{n+1} - b_n
\enq
 which correspond to ''physical'' initial data  for
the displacement field of the form
\beq
\label{2_4}
u_n (t=0) = a_n,\qquad \du_n (t=0) = b_n
\enq

Note that (\ref{2_2}), (\ref{2_3}) can be (formally)
rewritten  in the integral form
\bee
\label{2_5}
\al_n (t) = \be_n (t) + \chi_p \sum^\infty_{m=-\infty} \intto G_1
\left ( n - m, t - s \right ) \; \al^p_m (s) ds
\ene
\bee
\label{2_6}
\al_n (t) = \be_n (t) + \chi_p \sum^\infty_{m=-\infty} \intto G
\left ( n - m, t - s \right ) \; \De \al^p_m (s) ds
\ene
where $\be_n (t)$ is a solution of the Cauchy problem
\bee
\label{2_7}
\left\{
\begin{array}{l}
\ddbe_n (t) = \De \be_n (t) \\
\be_n (t=0) = \wta_n; \qquad \dbe_n (t=0) = \wtb_n\\
\lim_{n\to\infty} \be_n (t) = 0
\end{array}
\right.
\ene
\beq
\label{2_8}
G (n, t) = \frac{2}{\pi} \int^\pi_{-\pi} \cos\,(2\sig\;n) \;
\frac{\sin \,(2t\sin\;\sig)}{\sin\;\sig}\; d\sig,
\enq
is the Green function associated with (\ref{2_7}) and
\beq
\label{2_9}
G_1 (n, t) = \frac{2}{\pi} \int^\pi_{-\pi} \cos\, (2\sig n)
\sin\, (2t \sin\; \sig) \, \sin\;\sig \, d\sig,
\enq
with $n = 0, \pm 1, \pm 2, \cdots$.  We recall some elementary
properties of the functions $G_1 (n, t)$ and $G (n, t)$ which we will
use in what follows
\beq
\label{2_10}
|G_1 (0, t)| \le 8\, |t|,
\enq
\beq
\label{2_11}
|G_1 (n, t)| \le C \; \frac{|t|+|t|^3}{n^2},\qquad \forall n\ne 0,\;
\forall t\in\crr.
\enq
\beq
\label{2_12}
|\dot G (0, t)| \le 8,
\enq
\beq
\label{2_13}
|\dot G (n, t)| \le C \; \frac{|t|^2}{n^2},\qquad \forall n\ne 0,\;
\forall t\in\crr.
\enq
Hereafter various positive constants will be denoted as $C$ but,
we remark that they may vary from line to line.

Let $T>0$.  We consider the linear space $H = H (T)$ (it will be
refered to as the energy space) which consists of all functions
$\al (t)$ of the form:
\[
\al (t) = \left( \cdots, \al_{j-1} (t), \al_j (t), \al_{j+1} (t),
\cdots \right)
\]
such that
\[\mbox{(a)} \qquad \mbox{each} \qquad \al_j (t) \in C^2 ([0, T);\, \crr)\]
\[ \mbox{(b)} \qquad
\sup_{0\le t < T} \left[ \sum^{+\infty}_{n=-\infty} \left(
\sum^{n-1}_{m=-\infty} \dal_m (t) \right)^2 + \sum^{+\infty}_{n=-\infty}
\al^2_n (t) \right] < +\infty.
\]
We define the norm in $H$ as
\beq
\label{2_14}
||\al (\cdot)||^2_H = \sup_{0\le t < T} \left [ \sum^{+\infty}_{n=-\infty}
\left ( \sum^{n-1}_{m=-\infty} \dal_m (t) \right )^2 +
\sum^{+\infty}_{n=-\infty} \al^2_n (t) \right ]
\enq
and $(H, ||\cdot||_H)$ becomes a Banach space.  Thus $H$ is the space
of spatially localized functions $\al_n (t)$ having finite energy.
The respective localization law is implicitly defined by (\ref{2_14}).

\begin{lemma}
{\it Let $\be (t) = (\cdots,\be_{j-1}(t), \be_j (t), \be_{j+1} (t),
\cdots)$ where $\be_n (t)$ is the solution of the Cauchy problem
(\ref{2_7}).
Assume that $\{ \wta_n \}$ and $\{ \wtb_n \}$ satisfy the condition
\beq
\label{2_15}
\sum^{+\infty}_{n=-\infty} \left ( \sum^{n-1}_{m=-\infty} |\wtb_m|^2
\right)^2 + \sum^{+\infty}_{n=-\infty} |\wta_n|^2 < \infty.
\enq
Then, for any $T>0$ the function $\be (t)$ belongs to the energy space
$H=H(T)$}
\end{lemma}

{\it Proof}:
Lemma~1 follows from the explicit representation of the solution
$\be_n (t)$
\[
\be_n (t) = \sum^{+\infty}_{m=-\infty} \left[ \dot{G}(n - m, t) \,
\wta_m + G(n - m, t) \, \wtb_m \right]
\]
and the properties of the Green function. \qquad $\cx$

\bigskip
Let $R > 0$.  We define the subset
\[
F_R = \left \{ \al (t) \in H, \quad ||\al (\cdot) - \be (\cdot)||_H
\le R;\quad \al_n (0) = \be_n (0),\quad \dal_n (0) =
\dbe_n (0), \;\; \forall n \right \} \subseteq H.
\]
Clearly, $F_R$ is a closed subset of $H$.

\begin{theor} (Local Existence).
{\it Assume that $\{ \wta_n \}$ and $\{ \wtb_n \}$ are as in
Lemma~1.  Then, there exists $T>0$ and a unique function
\beq
\label{2_16}
\al (t) = \left( \cdots, \al_{j-1} (t), \al_j (t), \al_{j+1} (t),
\cdots \right)
\enq
defined in $[0, T)$, such that $\al_j (t)$ is a solution of
(\ref{2_2}), (\ref{2_3}),
and $\al (t)$ belongs to the Banach space $H$.}
\end{theor}

{\it Proof}:
Let us consider the map
\beq
\label{2_17}
P\al (t) = \left( \cdots, \wtpm \al_{j-1}(t), \wtpm \al_j (t),
\wtpm \al_{j+1} (t), \cdots \right)
\enq
where
\bee
\wtpm \al_n = \be_n (t) + \chi_p \sum^{+\infty}_{m=-\infty} \intto
G_1 (n-m, t-s) \, \al^p_m (s) ds =\nonumber \\
= \be_n (t) + \chi_p \sum^{+\infty}_{m=-\infty} \intto
G (n - m, t - s) \, \De\al^p_m (s) ds,\label{2_18} \\
\nonumber \qquad p \in {\crz}^+,\;\; p\ge 2,\;\; 0 \le t < T.
\ene
It is useful to note that
\beq
\label{2_19}
\wtpm \al_n (t=0) = \wta_n,\qquad
\fdt\; \wtpm \al_n (t=0) = \wtb_n.
\enq
The proof consists of two steps: (a) First, we show that $P:\; F_R
\mapsto F_R$ if $T$ is chosen sufficiently small.  (b) For the
second step we  show that $P$ is contraction on $F_R$, which means
that there exists $\ga$, $0 < \ga < 1$, such that
\[
||P f - P g||_H \le \ga ||f - g||_H;\qquad \forall f, g \in F_R
\]
provided $T$ is chosen sufficiently small.

Due to Lemma~1, in order to show that $P\al \in H$ whenever
$\al \in H$ it is sufficient to show that $P\al - \be \in H$.  To
this end we consider the two terms in the energy norm separately
and start with the ``potential energy''.  It follows from (\ref{2_18}) that
\beq
\label{2_20}
\sum^{+\infty}_{n=-\infty} \left [\wtpm \al_n (t) - \be_n (t)
\right ]^2 = \chi^2_p (A_1 + A_2 + A_3)
\enq
where
\bee
A_1 = \sum^{+\infty}_{n=-\infty}\sum^{+\infty}_{m=-\infty \atop m\ne n}
\sum^{+\infty}_{\ell=-\infty \atop \ell\ne n} \intto\!\! G_1 (n - m, t - s)\, \al^p_m (s) ds \intto \!\!G_1 (n - \ell, t - s_1) \, \al^p_\ell
(s_1)\, ds_1
\ene
\bee
\label{2_21}
A_2 = 2\sum^{+\infty}_{n=-\infty}\sum^{+\infty}_{m=-\infty\atop m\ne n}
\intto G_1 (n - m, t - s)\, \al^p_m (s) ds \intto G_1 (0, t - s_1) \,
\al^p_n (s_1)\, ds_1
\ene
\bee
\label{2_22}
A_3 = \sum^{+\infty}_{n=-\infty} \left ( \intto G_1 (0, t - s)
\, \al^p_n (s) ds \right )^2.
\ene
Using the fact that
\beq
\label{2_23}
\sum^{+\infty}_{n=-\infty} |\al_n (s)|^p |\al_n (s_1)|^p \le ||\al||^{2p}_H
\enq
it is straightforward to obtain that
\beq
\label{2_24}
\sup_{0\le t< T} |A_3| \le C ||\al||^{2p}_H \, T^4.
\enq
In order to estimate $|A_1|$ we first use (\ref{2_11}), what yields
\beq
\label{2_25}
|A_1| \le C \sum^{+\infty}_{n=-\infty} \left ( \sum_{k\ne 0} \intto
\frac{|t-s|+|t-s|^3}{k^2}\; |\al^p_{n-k} (s)|\, ds \right )^2.
\enq
Next, we take into account that $p\ge 2$ and employ (\ref{2_23}) to get
\beq
\label{2_26}
\sup_{0\le t < T} |A_1| \le C ||\al||^{2p}_H \, (T^2 + T^4)^2.
\enq
Similar calculations  to the above ones lead to
\beq
\label{2_27}
\sup_{0\le t < T} |A_2| \le C ||\al||^{2p}_H \, (T^2 + T^4)\, T^2.
\enq
Combining (\ref{2_24})-(\ref{2_27}) we get the estimate for
 the ``potential energy''
\beq
\label{2_28}
\sup_{0\le t < T} \sum^{+\infty}_{n=-\infty} \left [ \wtpm \al_n (t)
- \be_n (t) \right ]^2 \le C ||\al||^{2p}_H \, (T^2 + T^4)^2.
\enq

Next we calculate
\beq
\label{2_29}
\fdt\; \wtpm \al_n (t) - \dbe_n (t) = \sum^{+\infty}_{m=-\infty} \intto
\dot G (m, t - s)\,\De \al^p_{m+n} (s) ds.
\enq
It is easy to verify that the relation
\bee
\label{2_30}
\sum^{n-1}_{m=-\infty} \sum^{+\infty}_{\ell=-\infty} \dot G (m - \ell,
t - s) \, \De \al^p_\ell (s) = \nonumber \\
\qquad = \sum^{+\infty}_{\ell=-\infty} \left[ \dot G (m - \ell, t - s) -
\dot G (n - 1 - \ell, t - s) \right ]\, \left[ \al^p_{\ell+1} (s) -
\al^p_\ell (s) \right]
\ene
holds.  Consequently, we only need to estimate
\bee
\label{2_31}
\sum^{+\infty}_{n=-\infty} \left ( \sum^{+\infty}_{\ell=-\infty} \intto
\dot G (n - \ell - 1, t - s)\left [ \al^p_{\ell+1} (s) - \al^p_\ell
(s) \right ] ds \right )^2 \le B_1 + B_2
\ene
where
\[
B_1 = 2 \sum^{+\infty}_{n=-\infty} \left ( \sum^{+\infty}_{\ell\ne n-1}
\intto \dot G (n - \ell - 1, t - s) \left [ \al^p_{\ell+1} (s) -
\al^p_\ell (s) \right ] \, ds \right )^2
\]
and
\[
B_2 = 2 \sum^\infty_{n=-\infty} \left ( \intto \dot G (0, t - s)
\left [ \al^p_n (s) - \al^p_{n-1} (s) \right ] ds \right )^2.
\]

Using (\ref{2_12}), (\ref{2_13}) we estimate $B_1$ and $B_2$ as follows
\beq
\label{2_32}
|B_2| \le CT^2 ||\al (\cdot)||^{2p}_H
\enq
and
\beq
\label{2_33}
|B_1| \le CT^6 ||\al (\cdot)||^{2P}_H.
\enq

Finally, we have that
\bee
||P\al - \be||^2_H \le C ||\al||^{2p}_H \left ( T^4 + T^6 + T^8
\right) \nonumber \\
\le C (R+||\beta(0)||)^{2p}_H \left ( T^4 + T^6 + T^8
\right) \le R
\label{2_34}
\ene
if $T$ is chosen sufficiently small.  Thus, $P:\; F_R \mapsto F_R$
which shows (a).

Now we show that $P$ is a contraction map from $F_R$ into $F_R$
if $T$ is chosen small enough. Consider $\al (t)$, $\de (t) \in H$.
We can write
\bee
\label{2_35}
\sup_{0\le t < T} \sum^{+\infty}_{n=-\infty} \left [ \wtpm \al_n
(t) - \wtpm \de_n (t) \right ]^2 = \chi^2_p (C_1 + C_2 + C_3)
\ene
where
\bee
\label{2_36}
C_1= \sup_{0\le t < T} \sum^{+\infty}_{n=-\infty} \left ( \intto G_1 (0, t - s) \left [ \al_n (s) - \de_n (s) \right ] \left [
\al_n^{p-1} (s) + \cdots + \de_n^{p-1} (s) \right ]ds\right )^2
\ene
\bee
\label{2_37}
C_2= \sup_{0\le t < T} \sum^{+\infty}_{n=-\infty} \left(
 \intto G_1 (0, t - s) \left [ \al_n (s) - \de_n (s) \right ]\left[
\al_n^{p-1} (s) + \cdots + \de_n^{p-1} (s) \right ]ds \right )\times
\nonumber \\
\times \left ( \sum^{+\infty}_{\ell=-\infty \atop \ell\ne n} \intto G_1 (\ell - n, t - s_1) \left [ \al_\ell (s_1) - \de_\ell (s_1)
\right ]\left [ \al_\ell^{p-1} (s_1) + \cdots + \de_\ell^{p-1}
(s_1) \right ]ds_1 \right )
\ene
and
\bee
\label{2_38}
C_3 = \sup_{0\le t< T} \sum^{+\infty}_{n=-\infty} \left [ \intto
 \sum^{+\infty}_{m=-\infty\atop m\ne n} G_1 (n - m, t - s)\left [
\al_m (s) - \de_m (s) \right ]\right.\times\nonumber \\
\times \left. \left [ \al_m^{p-1} (s) + \cdots + \de_m^{p-1} (s)
\right ]ds \right]^2.
\ene
Straightforward estimates using (\ref{2_10}), (\ref{2_11})
 and similar to those provided above yield
\bee
\label{2_39}
|C_1| \le C ||\al (\cdot) - \de (\cdot)||^2_H T^4 \left ( ||\al(\cdot)||_H
+ ||\de(\cdot)||_H \right )^{2(p-1)}\nonumber \\
\le C ||\al (\cdot) - \de (\cdot)||^2_H T^4 R^{2(p-1)}
\ene
\bee
\label{2_40}
|C_2| \le C ||\al (\cdot) - \de (\cdot)||^2_H (T^4 + T^5) \,
(2 ||\beta(0) ||_H+2R)^{2(p-1)}
\ene
\bee
\label{2_41}
|C_3| \le C ||\al (\cdot) - \de (\cdot)||^2_H (T^5 + T^6) \,
(2 ||\beta(0) ||_H+2R)^{2(p-1)}
\ene
Therefore
\bee
\label{2_42}
\sup_{0\le t < T} \sum^{+\infty}_{n=-\infty} \left [ \wtpm \al_n (t)
- \wtpm \de_n (t) \right ]^2 \leq \nonumber \\ \leq
C(2 ||\beta(0) ||_H+2R)^{2(p-1)}
(T^4 + T^5 + T^6)
||\al (\cdot) - \de (\cdot)||^2_H  \leq \nonumber \\ \leq
C ||\al (\cdot) - \de (\cdot) ||^2_H.
\ene

Now we can estimate the ``kinetic'' term as follows: Let us write
\bee
\label{2_43}
\sup_{0\le t < T} \sum^{+\infty}_{n=-\infty} \left ( \sum^{+\infty}_{m=-\infty} \fdt \; \left [ \wtpm \al_m (t) - \wtpm \de_m (t) \right ]
\right )^2 = \chi^2_p (D_1 + D_2 + D_3)
\ene
with
\bee
\label{2_44}
D_1 = \sup_{0\le t < T} \sum^{+\infty}_{n=-\infty} \left ( \intto
\dot G (0, t - s) \left [ \al^p_n (s) - \de^p_n (s) - \al^p_{n-1}
(s) + \de^p_{n-1} (s) \right ]\, ds \right )^2
\ene
\bee
\label{2_45}
D_2 &= \sup_{0\le t < T} \sum^{+\infty}_{n=-\infty} \left ( \intto
G (0, t - s) \left [ \al^p_n (s) - \de^p_n (s) - \al^p_{n-1} (s) +
\de^p_{n-1} (s) \right ]ds \right ) \times\nonumber \\
&\times \left ( \sum^{+\infty}_{\ell=-\infty \atop \ell\ne n} \intto
\dot G (\ell, t - s_1) \left [ \al^p_{n+\ell} (s_1) - \de^p_{n+\ell}
(s_1) + \de^p_{n+\ell-1} (s_1) \right ]\, ds_1 \right )
\ene
and
\bee
\label{2_46}
D_3 = \sup_{0\le t< T} \sum^{+\infty}_{n=-\infty} \left [
\sum^{+\infty}_{\ell=-\infty\atop \ell\ne n} \intto \dot G (\ell, t - s)
\left [ \al^p_{n+\ell} (s) - \de^p_{n+\ell} (s) - \al_{n+\ell-1}^p (s)
\right ] \, ds \right ]^2.
\ene
In the same way as (\ref{2_39})--(\ref{2_41}) have been obtained
 but now using (\ref{2_12}) and (\ref{e2}) we get
\bee
\label{2_47}
D_1 &\le C ||\al (\cdot) - \de (\cdot)||^2_H \,
 (2 ||\beta(0) ||_H+2R)^{2(p-1)}\\
\label{2_48}
D_2 &\le C ||\al (\cdot) - \de (\cdot)||^2_H \, (T^2 + T^3)\,
(2 ||\beta(0) ||_H+2R)^{2(p-1)}\\
\label{2_49}
D_3 &\le C ||\al (\cdot) - \de (\cdot)||^2_H \, (T^3 + T^4)\,
(2 ||\beta(0) ||_H+2R)^{2(p-1)}
\ene
Combining the last three inequalities with (\ref{2_43}) and taking
$T > 0$ small enough so that
\[
C(2 ||\beta(0) ||_H+2R)^{2 (p-1)} (T^2 + T^3 + T^4 + T^5 + T^6) < 1
\]
then $P$ is a contraction in $F_R$.  Consequently, we have a unique
fixed point $\wtal \in F_R$, i.e., $P\wtal = \wtal$.  This means
that
\beq
\label{2_50}
\wtpm \wtal_j (t) = \wtal_j (t),\qquad \forall t \in [0, T)
\enq
i.e.
\beq
\label{2_51}
\wtal_j (t) = \be_j (t) + \sum^{+\infty}_{m=-\infty} \intto G_1
(j - m, t - s)\, \wtal^p_m (s) ds.
\enq
where $T > 0$ is chosen as above.\qquad $\cx$

{\bf Remark 1}.
We conclude this section mentioning that the norm we have chosen
above, having rather transparent physical meaning, is not the only
allowing us to prove the local existence.  We could have use for
example  a Banach space $X$ of functions
\[
\alpha(t)=(...,\alpha_{j-1}(t),\alpha_j(t),\alpha_{j+1}(t),...)
\]
such that $\alpha_j\in C^1([0,T);  I\!\!\!R)$ and
\beq
\label{2_52}
||\al (\cdot)||^2_X = \sup_{-\infty < n < \infty} (1 + n^2)
\left ( ||\al_n (\cdot)||^2_\infty + \lV \sum^{n-1}_{m=-\infty}
\dal_m (\cdot)\rV^2_\infty \right )
\enq
where
\[
||f_n (\cdot)||_\infty = \sup_{0 \le t < T} |f_n (t)|
\]
and for each $n\in\crz$, $\al_n (t)$ satisfies
\[
\sup_{0\le t < T} |\al_n (t)|< \infty, \qquad
\sup_{0\le t < T} \left\vert \sum^{n-1}_{m=-\infty} \dal_m (t)
\right\vert < \infty.
\]

In $\al (t) \in X$ then
\[
\al^2_n (t) \le \frac{||\al (\cdot)||^2_X}{1+n^2},\quad
\left ( \sum^{n-1}_{m=-\infty} \dal_m (t) \right )^2 \le
\frac{||\al (\cdot)||^2_X}{1+n^2},\qquad \forall n,\;\;
\forall t\in [0, T)
\]
which implies that $||\al (\cdot)||^2_H \le C ||\al (\cdot)||^2_X$
where $\| \cdot \|_H$ was defined in (\ref{2_16}).

\section{Global solution}

In order to prove global existence we will use a conserved quantity
of the problem at hand.  In terms of the depending variables
$\al_n (t)$ one of the conservation quantities is   the
Hamiltonian $\cah_\ell + \cah_{nl}$ itself. In fact, let
\beq
\label{3_1}
E(t)=\sum^{+\infty}_{n=-\infty} \left \{ \fud \,\left(
 \sum^{n-1}_{m=-\infty}
\dal_m (t) \right )^2 + \fud\; \al^2_n (t) + \frac{\chi_p}{p+1}\;
\al_n^{p+1} (t) \right \}
\enq
where $\alpha_m(t)$ is the solution found in Theorem 1. If we assume
that
\[
\sum^{+\infty} _{n=-\infty}|\tilde{a}_n|^{p+1} <+\infty
\]
then we can show that $E(t)$ is finite for any $t$ in the interval of
existence of $\alpha_n(t)$.
Straightforward calculation using the fact that $\alpha_n(t)$ solves
(\ref{2_2})-(\ref{2_3}) shows that $\frac{dE}{dt}=0$. Consequently,
$E(t)$=constant.

In order to discuss consequences of the conservation of the energy
we first notice that Theorem~1 gives the local existence of localized
excitations in FPU lattice in terms of the relative displacements
$\al_n (t)$.  Then, the question on the existence of the solution
in terms of original displacements of the particles arises.  It is
resolved by the following lemma:

\begin{lemma}
{\it Let $\{ \wtal_n \}$ and $\{ \wtb_n \}$ be as in Lemma~1 and
$\al = (\cdots, \al_n (t), \cdots)$ be the finite energy solution of
the Cauchy problem (\ref{2_2}), (\ref{2_3}) with $p\ge 2$.  Then, in the interval of
existence of $\al_n (t)$ the function
\beq
\label{3_2}
\wtu_n (t) = \sum^n_{m=-\infty} \al_m (t)
\enq
is well defined and it is bounded.}
\end{lemma}

{\it Proof}: Substitution of $\alpha_m$ by the expression in
(\ref{2_5}), using
the fact that $\sum^{+\infty}_{n=-\infty} |\al_n (t)|^{p+1}
< +\infty$
with $p\ge 2$ and $\al\in H$ for\break $t\in [0, T)$ proves the lemma.
\qquad $\cx$

\bigskip
The solution of the FPU lattice is obtained from the trivial
relation
\[
u_n (t) = \wtu_n (t) + \mbox{const}
\]

We know that Zorn's lemma \cite{Hays3} implies that
the local solution $\al (t)$
we found above can be extended to the maximal interval of existence
$0 \le t < T_{\max}$.  We want to show that $T_{\max} =+\infty$.

\begin{theor}
(Global Existence).
{\it Let $p\ge 3$ be an odd integer and $\chi_p > 0$.  Suppose that
the initial conditions for problem (\ref{2_2}), (\ref{2_3})
satisfy the assumptions of Lemma~1 and
\beq
\label{3_3}
\sum^{+\infty}_{n=-\infty} \tilde{a}_n^{p+1} < +\infty.
\enq
Then the finite energy solution $\al (t)$ exists globally and it
is unique.}
\end{theor}

{\it Proof}:
We consider an interval $0 \le T < T_{\max}$ where $T$ could be
very near to $T_{\max}$.  Let
\[
\al (t) = \left( \cdots, \al_{j-1} (t), \al_j (t), \al_{j+1} (t), \cdots
\right)
\]
be the solution of (\ref{2_2}), (\ref{2_3}) in the norm of $H$ given
in (\ref{2_16}).

Global existence in the energy norm follows as a consequence of the
conservation of $E(t)$ given in (\ref{3_1}), the fact that $p$ is an
odd integer greater than or equal to $3$ and $\chi_p > 0$.

It remains to show that the global solution in unique.
To this end suppose that
$\al (t)$ and $\de (t)$ are two solutions of (\ref{2_2}), (\ref{2_3})
with the same initial conditions at $t = 0$.  Let $T>0$ be fixed
 but arbitrary. Introduce
\[
v_n (t) = \sum^{n-1}_{m=-\infty} \al_m (t), \qquad
w_n (t) = \sum^{n-1}_{m=-\infty} \de_m (t).
\]
Due to Lemma~2, $v_n (t)$ and $w_n (t)$ are well defined and bounded
in any interval.  Clearly, $v_n (0) = w_n (0)$ and $\dot v_n (0) =
\dot w_n (0)$.  We claim that
\[
z_n (t) = v_n (t) - w_n (t) \equiv 0, \qquad \forall 0 \le t \le T.
\]
Denote by
\[
f_n (t) = \left [ v_n (t) - v_{n-1} (t) \right ]^{p-1} + \cdots +
\left [ w_n (t) - w_{n-1} (t) \right ]^{p-1}.
\]
Then equation for $z (t)$ takes the form
\bee
\label{3_4}
\left \{
\begin{array}{l}
\ddot z_n - \De z_n = \chi_p \left [ (z_{n+1} - z_n)\, f_{n+1}
- (z_n - z_{n-1})\, f_n \right ]\\
z_n (t = 0) = \dot z (t = 0) = 0.
\end{array}
\right.
\ene
Clearly the following relations
\bee
\label{3_5}
\fud\; \fdt \, \sum^\infty_{n=-\infty} \left \{ \dot z^2_n (t) +
\left [ z_{n+1} (t) - z_n (t) \right ]^2 \right \} = \nonumber \\
=  \chi_p \sum^\infty_{n=-\infty}
 \dot z_n (t) \left [ (z_{n+1} - z_n)\, f_{n+1} - (z_n - z_{n-1}) \, f_n
\right ].
\ene
holdds and using  Cauchy-Schwarz's inequality we deduce that
\bee
\left\vert \sum^\infty_{n=-\infty} \dot z_n (t) \left [ (z_{n+1}
- z_n)\, f_{n+1} - (z_n - z_{n-1})\, f_n \right ] \right\vert
\le
\nonumber \\
\le 2 \sup_{-\infty < n < \infty} |f_n (t)|
\left ( \sum^\infty_{n=-\infty} \right)^{\fud}
\left( \sum^\infty_{n=-\infty} (z_{n+1} - z_n)^2 \right )^{\fud}.
\nonumber
\ene
Next, we take into account that
\[
\sup_{-\infty < n < \infty} ||\al_n (\cdot)||_\infty^{p-1} \le
||\al (\cdot)||_H^{p-1},
\]
which implies that $\sup_{-\infty < n < \infty} |w_n| < \infty$.
Returning to the relation (\ref{3_5})  we obtain
that
\beq
\label{3_6}
\fdt \; \varphi (t) \le C\; \sup_{-\infty < n < \infty} |f_n (t)|
\; \varphi (t) \le C_1 \varphi (t)
\enq
where
\[
\varphi (t) = \sum^\infty_{n=-\infty} \left \{ \dot z^2_n (t) +
\left [ z_{n+1} (t) - z_n (t) \right ]^2 \right \} \ge 0.
\]
From (\ref{3_6}) and Gronwall's lemma we deduce that
$\varphi (t)$ is identically zero.\qquad $\cx$

{\bf Remark 2}. The FPU lattice considered in this section if of
great practical importance and that is why received much attention in
physical literature where it has been treated with help of numerous
approximations. One of the approaches,  the so-called long-wavelength
limit (or continuum limit) is often used in order to reduce the
system of differential-difference equations to an evolution partial
differential equation, namely the generalized Korteweg-de Vries
(GKdV) equation
\bee
\label{3_7}
\frac{\p v}{\p t} + \frac{\p^3 v}{\p x^3} + \chi_p v^{p-1}
\; \frac{\p v}{\p x} = 0.
\ene
with $-\infty<x<+\infty$, $t>0$. It is well known that (\ref{3_7})
for $p$ large enough displays a dispersive blow up. This was shown by
Bona and Saut in \cite{BS} in special situations but in general, for
$p>5$ there is only numerical evidence that blow up really happens.
We remark that this difficulty does not occur in the discrete case
treated above which allows us to study strong interactions.

\section{Some other important models}

In this section we shall consider some additional models for which the
global existence (and uniqueness) of the solution can be found in the
same way we proceed in the previous sections. First, we consider the
sine-Gordon lattice with nonlinear inter-site interactions
\beq
{\cal H}_{nl}=K\sum_{n=-\infty}^{\infty}[1-\cos(u_{n+1}-u_{n})]
\enq
and linear part is given by (\ref{e2}). The dynamical equations in
terms of the relative displacements $\alpha_n(t)$ read
\beq
\label{4_1}
\ddal_n - \De \al_n - \chi \De \sin \; \al_n = 0.
\enq
The only difference compared with (\ref{2_2}) consists on  type
of nonlinearity.  Hence we can continue using the space $H$
we considered in Section 2 with the norm (\ref{2_14}) and use
Lemma~1.  Since $|\sin\; f| \le |f|$ one reduces
the problem at hand to the considerations given in Theorem~1 to obtain
local solutions. In order to obtain global existence we consider the
function
\[
E(t)=\sum^\infty_{n=-\infty} \left \{\fud \left ( \sum^{n-1}_{m=-\infty}
\dal_m (t) \right )^2 + \fud \; \al^2_n (t) + \chi [1 - \cos\;
\al_n (t) ] \right\}
\]
Using (\ref{4_1}) it is eassy to prove that $\frac{dE}{dt}=0$.
Therefore
\beq
\label{4_2}
E(t)=\mbox{constant}=E(0)
\enq

Thus, whenever the constant $\chi$ is positive then,  (\ref{4_2}) will
tell us that the solution
$\alpha(t)=(...,\alpha_{j-1}(t),\alpha_j(t),\alpha_{j+1}(t),...)$
remains bounded in the norm of the space $H$, which proves
that the solution exist globally.
Uniqueness can be shown as we proceeded in Theorem~2.

Next we consider the so-called a lattice with on-site nonlinearity. In
this case ${\cal H}_l$ is as in (\ref{e2}) and
\[
\cah_{n\ell} = \cah_{OS} = \frac{K_{p+1}}{p+1} \sum^\infty_{n=-\infty}
u_n^{p+1}
\]
The dynamical equations read
\beq
\label{4_3}
\ddot u_n - \De u_n + \chi_p \; u^p_n = 0
\enq
In this case  it is convenient to explore again the integral form
(\ref{2_7}). To prove local existence we define for $t>0$ the linear
space $\tilde{H}=\tilde{H}(T)$ which consists of all functions
\[
u (t) = (\cdots, u_{j-1} (t), u_j (t), u_{j+1} (t), \cdots)
\]
such that
\[ \mbox{(a)} \qquad u_j (t) \in C^2 \left([0, T_0);\;
\crr\right)
\]
and
\[
\mbox{(b)}  \qquad
\sup_{0\le t < T} \sum^\infty_{n=-\infty} \left [ \dot u^2_n (t)
+ u^2_n (t) \right ] < +\infty.
\]

The energy norm in $\tilde{H}$  is defined as
\[
||u (t)||^2_{\tilde{H}} = \sup_{0\le t < T} \left \{ \sum^\infty_{n=-\infty}
\left [ \dot u^2_n (t) + u^2_n (t) \right ] \right \}.
\]
and $(\tilde{H}, \|\cdot\|_{\tilde{H}})$ is a Banach space. we can
easily show that if
   \[
\be (t) = \left(\cdots, \be_{j-1} (t), \be_j (t), \be_{j+1}
(t), \cdots\right)
\]
 where $\be_j (t)$ is the solution of  (\ref{2_9}) with
\beq
\label{ee1}
\sum^\infty_{n=-\infty} \left( |\tilde{a}_n|^2 + |\tilde{b}_n|^2
 \right) < +\infty
\enq
then, the function $\be (t)$ belongs to the space $\tilde{H}$ for any
$T>0$.

Under the above assumptions we can prove local existence of (\ref{4_3})
if $T$ is chosen sufficiently small. Furthermore, if we assume
additionally that
\[
\sum_{n=-\infty}^{+\infty}|\tilde{a}_n|^{p+1}<+\infty
\]
we can  consider the function
\beq
\label{4_4}
E(t)=\frac 12
\sum_{n=-\infty}^{+\infty}[\dot{u}_n^2(t)+u_n^2(t)]+
\frac{\chi_p}{p+1}\sum^{+\infty}_{n=-\infty}u_n^{p+1}(t)
\enq
and show that $E(t)$ is finite for any $t$ in the interval of
existence. Using (\ref{4_3}) we can prove that $\frac{dE}{dt}=0$,
therefore $E(t)=$constant$=E(0)$ for any $t$ in the interval of
existence. hence, if we assume that $\chi_p>0$  and $p=$odd$\geq 3$
we obtain the desired a priori estimate
for the norm $\| u(\cdot)\|_{\tilde{H}}$ which proves global existence.
 Uniqueness can be shown as in Theorem 2. Similar
discussions can be done for the so-called sine-Gordon on-site
nonlinear lattice
\[
\ddot{u}_n-\Delta{u}_n+\chi\sin u_n=0.
\]

\section{Acknowledgements}

The second author is grateful to Prof. S.~Rauch-Wojciechowski for a number of
useful comments.  The first author acknowledges the hospitaliy at the Department
of Physics (University of Madeira) while he was visiting and completing
this research.  The work has been supported by the bilateral program
JNICT/CNPq (Portugal/Brazil).  VVK acknowledges support from FEDER and
the program PRAXIS XXI, grant n$^o$ PRAXIS/2/2.1/FIS/176/94
and GPM was partially supported by a Grant of CNPq and PHONEX (MCT,
Brazil).

\end{document}